\documentclass[aps,pra,twocolumn,superscriptaddress,longbibliography,nobalancelastpage,floatfix]{revtex4-2}

\usepackage{graphicx} 
\usepackage[hypertexnames=false]{hyperref}
\usepackage{physics}
\usepackage{orcidlink}

\usepackage{comment}

\begin{document}
\title{Extracting Coupling-Mode Spectral Densities with Two-Dimensional Electronic Spectroscopy}
\author{Roosmarijn de Wit}
\affiliation{SUPA, School of Physics and Astronomy, University of St Andrews, St Andrews, KY16 9SS, United Kingdom}
\author{Jonathan Keeling}
\affiliation{SUPA, School of Physics and Astronomy, University of St Andrews, St Andrews, KY16 9SS, United Kingdom}
\author{Brendon W. Lovett}
\affiliation{SUPA, School of Physics and Astronomy, University of St Andrews, St Andrews, KY16 9SS, United Kingdom}
\author{Alex W. Chin}
\affiliation{Sorbonne Université, CNRS, Institut des NanoSciences de Paris, 4 place Jussieu, 75005 Paris, France}
\date{\today}

\begin{abstract}
    \textbf{Abstract:} Methods for reconstructing the spectral density of a vibrational environment from experimental data can yield key insights into the impact of the environment on molecular function. Although such experimental methods exist, they generally only access vibrational modes that couple diagonally to the electron system. Here we present a method for extracting the spectral density of modes that couple to the transition between electronic states, using two-dimensional electronic spectroscopy. To demonstrate this, we use a process-tensor method that can simulate two-dimensional electronic spectroscopy measurements in a numerically exact way. To explain how the extraction works, we also derive an approximate analytical solution, which illustrates that the non-Markovianity of the environment plays an essential role in the existence of the simulated signal.
\end{abstract}

\maketitle

Environment-mediated processes play an important role in the photophysics of molecules~\cite{mukamel:1995, chin:2013, caycedo:2022, gustin:2023, schroder:2019, womick:2011, arsenault:2020, sil:2022, moya:2022, hunter:2024, sohoni:2024}. The electron-vibrational interactions involved in such processes are often characterized by a spectral density that encodes the coupling strength to each mode in the environment~\cite{mukamel:1995, breuer:2002}. Determination of the spectral density is therefore key to gaining a full description of these interactions. Experimentally, this can be achieved with techniques such as resonance Raman and fluorescence line narrowing spectroscopy~\cite{mukamel:1995, pachon:2014, gustin:2023, ratsep:2007, ratsep:2008, ratsep:2009}. These methods are generally sensitive to vibrational modes that tune the electronic energy levels and thus couple diagonally to the electronic eigenstates. However, spectral densities that characterize the off-diagonally coupled modes are challenging to extract~\cite{gustin:2023}. These coupling modes are known to mediate processes such as singlet fission~\cite{schroder:2019}, intensity borrowing~\cite{yin:2020} and efficient excitation transfer in organic chromophores~\cite{womick:2011, arsenault:2020, sil:2022, moya:2022, sohoni:2024}. Therefore, the ability to directly extract the coupling-mode spectral density could enhance our understanding of such phenomena, and could provide the potential to leverage the environment to enhance molecular function.

In this Letter, we present a method for extracting the spectral density of coupling modes with two-dimensional electronic spectroscopy (2DES)~\cite{mukamel:1995, cho:2008, collini:2021, fresch:2023}. It relies on directly probing electronic transitions that are coupled vibrationally in the system of interest, resulting in a signal that encodes the spectral density of the coupling modes. We simulate this signal for a three-level chromophore system. To explain how our method works, we derive an approximate analytical solution to the signal, which restricts the vibrational manifold to single phonon interactions. We compare this solution to a numerically exact computation that employs tensor network techniques to capture the influence of the environment.

Using non-linear response theory, the detected signal in a 2DES experiment is usually expressed as a sum of four-time correlation functions~\cite{mukamel:1995, cho:2008}. For the 2DES pulse sequence explored in this work, we find that the corresponding multi-time correlation function only has a non-zero solution for a non-Markovian treatment. That is, in the case where the environment can retain a memory of past interactions with the system~\cite{breuer:2002}. In contrast, we find that the signal we study vanishes when applying standard master equation techniques which assume a Markovian environment. Our proposed method for extracting spectral densities may therefore also be relevant to the study of devising experimental protocols for probing non-Markovianity. 

Studying non-Markovianity in the context of multi-time correlation functions requires careful consideration.
Under a Markovian assumption, one can  use the quantum regression theorem to find multi-time correlations straightforwardly~\cite{breuer:2002}. The quantum regression theorem may not always give correct results however, even if the corresponding dynamical map accurately predicts single-time observables~\cite{guarnieri:2014}. In such cases, computational techniques that can efficiently describe non-Markovian environments are necessary~\cite{vega:2000, tanimura:2020, mascherpa:2020, chen:2022, gera:2023, sirkina:2023}. One particular approach is to express the open quantum system as a tensor network that effectively compresses the exponentially large Hilbert space to only the physically most relevant degrees of freedom~\cite{orus:2014, ng:2023, chin:2013, tamascelli:2019, somoza:2019}. This includes the method employed in this work, which relies on a process tensor (PT)~\cite{pollock:2018} that captures all possible effects of the environment on the system. By casting the process tensor into a matrix product operator (PT-MPO) format, computations can be performed in an efficient and numerically exact way~\cite{pollock:2018,jorgensen:2019, strathearn:2017, strathearn:2018, fux:2021, cygorek:2022, link:2023, cygorek:2024, oqupaper}. Moreover, since the PT-MPO maps any intermediate control operations onto the final state, this framework lends itself particularly well to the computation of multi-time correlations.

\begin{figure}
\includegraphics[width=8.5cm]{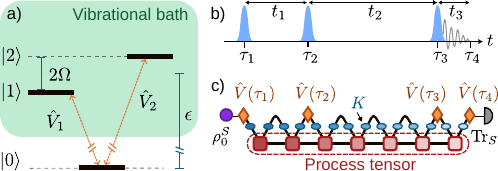}
\caption{a) Sketch of our model, consisting of a three-level system coupled to a bosonic bath. b) Laser pulses in a 2DES experiment, applied at times $\tau_{1,2,3}$. The signal is emitted at time $\tau_4$. Time intervals are denoted by $t_{1,2,3}$. c) Tensor network of seven time steps for calculating a four-time correlation function with a process tensor (red squares) and initial system state $\rho_0^S$ (purple circle). Each dipole operator $\hat{V}$ (orange diamonds) is applied at times $\tau_{i}$. Propagators related to the system Hamiltonian are denoted by $K$ (blue ovals).} 
\label{fig1}
\end{figure}

We start by introducing the model we are considering in this Letter, shown in Fig.~\ref{fig1}a). It involves a three-level electronic system, linearly coupled to a many-mode vibrational bath. This corresponds to a Hamiltonian of three components, $\hat{H} = \hat{H}_S+\hat{H}_B+\hat{H}_I$, which respectively describe the electronic system, vibrational bath and system-bath interaction:
\begin{multline}
    \hat{H} = (\epsilon - \Omega)\dyad{1}{1} + (\epsilon + \Omega) \dyad{2}{2} + \sum_k\,\omega_k b_k^{\dagger}b_k \\
    +  \pqty{\dyad{1}{2} +\mathrm{H.c.}}\sum_k(g_k b_k + g_k^{*}b_k^{\dagger})
    \label{ham}
\end{multline}

The average energy of the electronic eigenstates, $\ket{1}$ and $\ket{2}$, is set by $\epsilon$, which defines the overall electronic energy scale, while $\Omega$ defines their splitting. 
The vibrational modes of the bath with frequencies $\omega_k$ are described by creation (annihilation) operators $b^\dagger_k (b_k)$. The coupling of each mode to the system is set by $g_k$ and characterized by the spectral density  $J(\omega)~=~\sum_k\abs{g_k}^2\delta (\omega - \omega_k)$.

A model of this form can describe a range of organic chromophore systems, such as carotenoids~\cite{manawadu:2023} and methylene blue~\cite{dunnett2021influence}. The photophysics of such molecules are generally described by three relevant energy levels, $S_0$, $S_1$ and $S_2$, corresponding to the states $\ket{0}$, $\ket{1}$ and $\ket{2}$ in our model. Alternatively, $\ket{1}$ and $\ket{2}$ could represent the lowest excited states of two coupled chromophores in a single-excitation subspace, as found in photosynthetic complexes~\cite{womick:2011, sil:2022, arsenault:2020, oviedo:2010}. Organic chromophores generally absorb light in the visible frequency range, while the characteristic timescale of the electron dynamics is typically slower, on the order of femtoseconds to picoseconds~\cite{arsenault:2020, sil:2022, moya:2022}. We will therefore show our results in a rotating frame, by effectively setting $\epsilon$ = 0 in subsequent calculations. Additionally we sill set  $\Omega = 2$~ps$^{-1}$. We do not consider explicit coupling between the excited and ground state manifolds in the Hamiltonian;  
instead we will consider coupling between the ground and excited states via the externally applied pulses in a 2DES experiment, as discussed below.

The bath modes represent the intramolecular and solvent vibrations, with which the system can interact. Importantly, the coupling between the system and bath is of an off-diagonal form, meaning that the electronic excited states ($\ket{1},\ket{2}$) are connected via the coupling to the bath. To test our proposed extraction method, we will consider two different spectral densities. Firstly, we extract a spectral density of an Ohmic form:
\begin{equation}
    J_1(\omega) = 2 \alpha_1 \omega e^{-\frac{\omega}{\omega_{1}}},
    \label{sd}
\end{equation}
where $\alpha_1$ is a dimensionless parameter that sets the system-bath coupling strength and $\omega_{1}$ denotes the cutoff frequency of the bath. Ohmic spectral densities are commonly used in optical spectroscopy simulations, although realistic spectral densities for chromophore models often contain more structure~\cite{ratsep:2007, ratsep:2008, ratsep:2009, sohoni:2024, gustin:2023}. We therefore secondly investigate a spectral density with some additional structure, containing an Ohmic and super-Ohmic component:
\begin{equation}
    J_2(\omega) = 
    2 \alpha_2 \omega e^{-\frac{\omega}{\omega_{2}}} + 
    2 \alpha_3 \frac{\omega^3}{\omega^2_{c3}} e^{-\frac{\omega}{\omega_{3}}},
    \label{sd2}
\end{equation}
where $\alpha_{2,3}$ and $\omega_{2,3}$ are respectively the system-bath coupling parameters and the cutoff frequencies of the Ohmic and super-Ohmic contributions.

Having introduced the model, we now discuss how the coupling-mode spectral density may be extracted with a 2DES experiment.
In general, the response in a 2DES experiment consists of a number of interaction pathways, depending on the experimental phase-matching conditions~\cite{mukamel:1995, cho:2008, collini:2021, fresch:2023}. Using non-linear response theory, each pathway can be expressed as a four-time correlation function~\cite{mukamel:1995, cho:2008}. To extract the spectral densities in Eqs.~\eqref{sd} and~\eqref{sd2}, we employ a particular pulse sequence that outputs a signal corresponding to a correlation function of the form:
\begin{equation}
    R = \Tr\bqty{\hat{V}_2(\tau_4)\hat{V}_1(\tau_3)\hat{V}_1(\tau_2)\hat{V}_2(\tau_1)\rho_0}.
    \label{seq}
\end{equation}
As shown in Fig.~\ref{fig1}a), this pulse sequence involves two different dipole operators that probe two distinct electronic transitions: $\hat{V}_1=\ketbra{1}{0}+\mathrm{H.c.}$ and $\hat{V}_2=\ketbra{2}{0} +\mathrm{H.c.}$. The operators are evaluated at the times of laser pulse interactions and signal emission, denoted by $\tau_i$. We will use $t_i$ to refer to the time intervals between pulses, as shown in Fig.~\ref{fig1}b).  We will take the initial system+bath state $\rho_0$ as the tensor product of the system ground state $\ketbra{0}{0}$ and the thermal state of the bath at temperature $T=100$~K. 

We next give a brief explanation of why the pulse sequence in Eq.~\eqref{seq} allows for the extraction of the spectral density; this will be explored in more detail below.
The sequence starts at time $\tau_1$ by creating an optical coherence between the ground and second excited state $\ket{2}$. Due to the off-diagonal system-bath coupling in Eq.~\eqref{ham}, the electronic state can then only evolve into the first excited state manifold $\ket{1}$ by absorbing or emitting vibrational excitations (phonons). Probing the system with the second pulse ($\hat{V}_1$) produces a response that depends on such transitions, and hence encodes the system-bath entanglement accumulated during $t_1$. Measuring the linear response $\Tr\bqty{\hat{V}_1(\tau_2)\hat{V}_2(\tau_1)\rho_0}$ is however not sufficient to extract the coupling-mode spectral density, as this two-time correlation function exactly vanishes: this pulse sequence has transferred the initial thermal vibrational state to vibrational coherences, which disappear when tracing over the bath at this stage. The two further operations in the 2DES signal can reverse the creation of vibrational coherences, allowing for a non-zero signal.  The dependence of this signal on $t_2$ can then be used to directly probe the spectral density of the vibrational modes involved in the $\ket{1}\leftrightarrow\ket{2}$ transitions.

\begin{figure*}
\includegraphics[width=14cm]{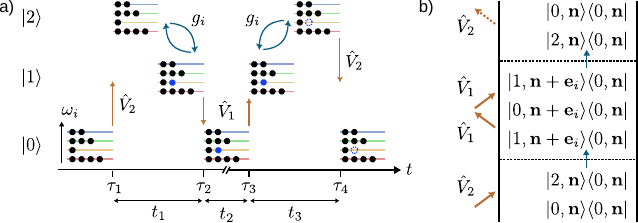}
\caption{a) Diagrammatic representation of one contribution to the system-bath evolution during the spectral density extraction protocol. Each horizontal line represents a vibrational bath mode, depicted  within the manifold of the relevant electronic state ($\ket{0}, \ket{1}, \ket{2}$). For simplicity, only four bath modes are drawn while the model contains a continuum of modes. The dots on each line show the number of phonons in a particular mode. The absorption/emission of a phonon in mode $i$ is denoted by a blue solid dot or a dashed outline respectively. The laser pulses and output signal are represented by orange arrows and the corresponding dipole operators. The double blue arrows represent the build-up of a superposition state in the excited states manifold, resulting from the system-bath coupling $g_i$. b) Double-sided Feynman diagram of the response in Eq.~\eqref{seq}. The evolution of the internal state during $t_1$ and $t_3$ is denoted by a dashed line and blue arrow. For clarity, only the bath states resulting from the initial absorption and eventual emission of a phonon in mode $i$ are shown in a) and b).} 
\label{fig2}
\end{figure*}

In order to compute the four-time correlation function $R$ in Eq.~\eqref{seq} in a numerically exact way, we use a PT-MPO  that can capture any non-Markovian effects on the system's evolution. In particular, we will use the process tensor-time evolving matrix product operator (PT-TEMPO) method, which constructs the PT-MPO from a path integral formalism~\cite{strathearn:2017, pollock:2018,  strathearn:2018, fux:2021, oqupaper, oqupy}. We provide a brief overview of the computation method here; for a more detailed description, we refer the reader to Refs.~\cite{oqupaper, dewit:2025}. 

Fig.~\ref{fig1}c) illustrates how multi-time correlations are calculated with our method. Because the evolution resulting from the bath and system-bath interaction in Eq.~\eqref{ham}) is encoded in the process tensor, the initial input state is the reduced system density matrix $\rho_0^S$ (purple circle). The tensor network is expressed in Liouville space, such that density matrices are represented by vectors and superoperators by matrices. $\rho_0^S$ is therefore drawn as a tensor with a single leg, while the dipole operators (orange diamonds) and system propagators $K$ (two per single time step, blue ovals) have two legs. The red squares represent the PT-MPO, drawn here for seven time steps. To calculate a multi-time correlation function, the system propagators and dipole operators are connected to the PT-MPO at the relevant time steps~\cite{gribben:2022, fux:2023, dewit:2025}. The bonds in the network are subsequently contracted, which corresponds to tracing over the system subspace (gray semi-circle). Crucially, the process tensor does not depend on system Hamiltonian or any measurement superoperators. It therefore only needs to be constructed once and can be re-used when varying the times at which the dipole operators are evaluated. We provide the convergence parameters and computational resources required in Section 1 of the Supporting Information.

To gain a deeper physical insight beyond the numerical computation, we next derive an approximate analytical expression for the signal in Eq.~\eqref{seq}. In contrast to a Markovian master equation, this derivation explicitly tracks the evolution of the (approximate) eigenstates of the full system+bath Hamiltonian in Eq.~\eqref{ham}, excited by the $\hat{V}_{1,2}$ operators, and does not assume factorisation between the system and bath at intermediate times. To do this, we restrict the bath Hilbert space to configurations in which only a single phonon is absorbed or emitted after each application of a dipole operator. We start from a particular configuration of the thermal ground state, 
\begin{equation}
    \ket{0, \vb{n}} = \ket{0}\otimes\ket{n_1,n_2, \ldots, n_i, \ldots},
    \label{init}
\end{equation}
where $n_i$ corresponds to the occupation of mode $i$. The initial density matrix of the system and bath is therefore $\rho_0 = \sum_{\vb{n}} P_{\vb{n}} \ket{0, \vb{n}} \bra{0, \vb{n}}$ where $P_{\vb{n}}\propto\exp(-\beta \sum_i \omega_i n_i)$ is the Boltzmann-weighted probability of the occupation vector $\vb{n}$.
We then apply the $\hat{V}_{1,2}$ operators at times $\tau_{1,2,3,4}$ according to Eq.~\eqref{seq} and track the time evolution of the excited states. This is diagrammatically depicted in Fig.~\ref{fig2}a), where we show the evolution of four bath modes in each electronic manifold, represented by the horizontal lines in each sub-diagram. The number of dots correspond to the number of phonon excitations in each mode and thus to an example state $\vb{n}$ drawn from the thermal configuration in $\rho_0$. From the form of the system-bath interaction in Eq.~\eqref{ham}, it follows that initial excitation of the ground state with $\hat{V}_2$ leads to the formation of a superposition in both excited state manifolds during $t_1$. The ket component of $\rho_0$ will be:
\begin{multline}
    \ket{2, \vb{n}(t_1)} \approx e^{-iE_{\vb{n}}t_1  }\Big(\alpha(t_1)\ket{2, \vb{n}}\\
    + \sum_i\bqty{\beta_{i}(t_1)\ket{1, \vb{n} + \vb{e}_i }+\gamma_{i}(t_1)\ket{1,\vb{n} -\vb{e}_i }}\Big),
    \label{plusn}
\end{multline}
neglecting higher order changes of the vibrational number state as noted above.  We have denoted the absorption/emission of a phonon in mode $i$ by the vector $\pm \vb{e}_i$, such that $[\vb{e}_i]_j, = \delta_{ij}$. The coefficients $\alpha$, $\beta_{i}$ and $\gamma_{i}$ evolve during the first time interval $t_1$ according to the time-dependent Schrödinger equation. Additionally, the phase arising from the total energy of the initial configuration $E_{\vb{n}} = \sum_i \omega_in_i$ has been factored out. Since the bra-side of the density matrix will evolve with exactly the conjugate of this phase, this term vanishes in the final expression. In Fig.~\ref{fig2}a), we illustrate one process contributing to $\ket{2, \vb{n}(t_1)}$, corresponding to the creation of a phonon in a single mode (additional blue dot), governed by the evolution of  $\beta_{i}$ in Eq.~\eqref{plusn}.

At $\tau_2$, we apply the $\hat{V}_1$ operator to obtain
\begin{equation}
    \hat{V}_1\ket{2, \vb{n}(t_1)} \approx \sum_i\bqty{\beta_{i}(t_1)\ket{0, \vb{n} + \vb{e}_i }+\gamma_{i}(t_1)\ket{0, \vb{n} - \vb{e}_i  }}.
    \label{plusn2}
\end{equation}
At this point, we have created a superposition of bath states in the electronic ground state that will evolve during $t_2$ according to the bath Hamiltonian:
\begin{multline}
  e^{-iE_{\vb{n}}(t_1 + t_2)  }\sum_i\Big[\beta_{i}(t_1) e^{-i\omega_it_2} \ket{0, \vb{n} + \vb{e}_i }\\
    +\gamma_{i}(t_1) e^{+i\omega_it_2} \ket{0, \vb{n} - \vb{e}_i  }\Big].
    \label{t2}
\end{multline}
Note here that if we would now take the trace to calculate the two-time correlation function  $\Tr\bqty{\hat{V}_1(\tau_2)\hat{V}_2(\tau_1)\rho_0}$, we would find a solution of zero, because the vibrational state in Eq.~\eqref{t2} is never $\ket{\vb{n}}$.

We now repeat the same procedure in reverse by applying $\hat{V}_1$ and $\hat{V}_2$ at times $\tau_3$ and $\tau_4$ respectively, described in detail in Section 2 of the Supporting Information. As shown in Fig.~\ref{fig2}a), the final two operators enable the back-flow of information from environment to system, generating a non-zero signal and allowing the extraction of the system-bath correlations at the final time $\tau_4$. We similarly find a set of coefficients that the describe the evolution of the system-bath states during $t_3$. Through solving the time-dependent Schrödinger equation and employing Weisskopf-Wigner theory, we obtain a set of closed equations of motion~\cite{weisskopf:1930, breuer:2002}. We show the determination of the coefficients as a function  of $t_1$ in detail in Section 3 of the Supporting Information. 
The final solution for $R$ is then found by using the definition of the spectral density and Boltzmann-weighted probabilities of occupations to give:
\begin{equation}
    R= \int_0^\infty \dd \omega \bqty{R_\beta(t_1,t_3,\omega) e^{-i\omega t_2} + R_\gamma(t_1,t_3,\omega) e^{i\omega t_2}},
    \label{r2}
\end{equation}
with
\begin{flalign}
        &R_\beta (t_1,t_3,\omega) =  J(\omega)\pqty{N(\omega)+1} f_{+\omega} (t_1) f_{+\omega} (t_3); \label{rbeta}\\
        &R_\gamma (t_1,t_3,\omega) = J(\omega)N(\omega)f_{-\omega} (t_1) f_{-\omega} (t_3);
        \label{rgamma}\\
        &f_{\pm\omega}(t) =  \frac{\pqty{e^{-\Gamma_2 t -iE_2^\prime t} - e^{-\Gamma_1t-i\pqty{\pm\omega+E_1^\prime}t}}  }{\pqty{E^\prime_2-E^\prime_1 - i\Gamma_1 + i\Gamma_2 \mp \omega}}.\label{fw} 
\end{flalign}
To obtain the above form of $f_{\pm\omega}(t)$, we have for simplicity assumed that the influence of the environment on the evolution of $\alpha(t_1)$ in Eq.~\eqref{plusn} is Markovian, and similar for the evolution during $t_3$. This results in the decay rates $\Gamma_{1,2}$ and Lamb shifts $S(\pm 2\Omega)$ that modify the system eigenvalues in Eq.~\eqref{ham}, specified in Section 3 of the Supporting Information. Here the Lamb shifts have been absorbed in the system eigenvalues, such that $E_1^\prime = \epsilon -\Omega + S(-2\Omega)$ and $E_2^\prime = \epsilon + \Omega + S(2\Omega)$.
Were one to use a fully Markovian approach to calculate $R$, this would neglect the system bath entanglement that exists during the time evolution.  Since, as discussed above, that entanglement is crucial to the measured signal, a fully Markovian approach cannot capture this effect.

\begin{figure}
\includegraphics[width=8.45cm]{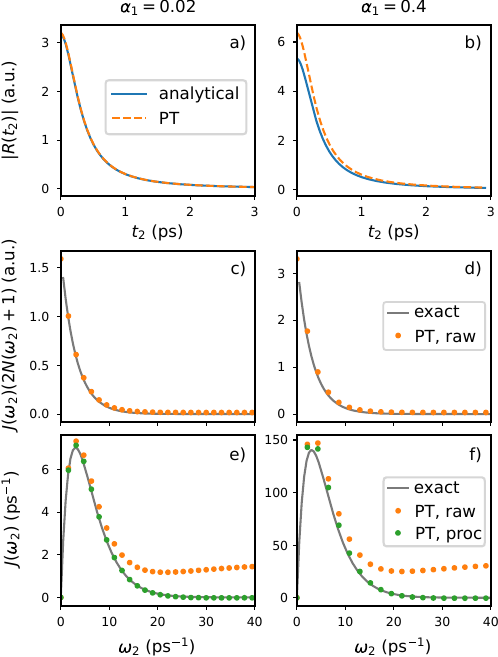}
\caption{Simulated response for an Ohmic spectral density, Eq.~\eqref{sd}, and two system-bath coupling strengths $\alpha_1$. a-b) The modulus of the response in the time domain as a function of $t_2$ and $t_1=t_3=0.01$~ps, calculated with the analytical solution in Eq.~\eqref{r2} (blue) and PT-TEMPO (orange dashed). c-d) Extracted bath information computed with PT-TEMPO (orange dots), compared to the expected result (grey line). The offset has not been subtracted from the results, discussed in the main text. e-f) The extracted spectral density computed with PT-TEMPO and compared to the expected result (grey line). The orange dots represent the raw output without the offset subtracted; green dots correspond to the processed data.} 
\label{fig3}
\end{figure}

Our final expression for the signal $R$ consists of two components, $R_\beta$ and $R_\gamma$, that represent the build-up of system-bath entanglement during the first and final time interval. $R_\beta$ encodes the creation and subsequent destruction of a phonon during $t_1$ and $t_3$ respectively, while $R_\gamma$ represents the reverse process. During the waiting time $t_2$, all excitations are contained within the bath, such that the density matrix evolves according to a simple phase (Eq.~\eqref{r2}). Inspired by beating maps which can be extracted from 2DES measurements~\cite{bakulin:2016, petropoulos:2024}, we can exploit this simple phase dependence to extract the spectral density by Fourier transforming over the waiting time:
\begin{equation}
    \Tilde{R}(t_1,\omega_2,t_3) = \int_0^\infty \dd \omega\int_0^\infty \dd t_2 e^{i \omega_2 t_2} \mathrm{Re}\bqty{{R(t_1,t_2,t_3)}}.
    \label{FT}
\end{equation}
Here we have chosen to Fourier transform only the real part of $R$, such that when we solve the transform using the relation 
\begin{equation}
    \int_0^\infty \dd t e^{\pm i \omega t} = \pi \delta(\omega) \pm i \mathcal{P} \frac{1}{\omega},
\end{equation}
we can discard the Cauchy principle value component, denoted by $\mathcal{P}$, by selecting the real part again after transforming. We then obtain:
\begin{equation}
    \Re \bqty{\Tilde{R}(t_1,\omega_2,t_3)} = \frac{\pi}{2} \pqty{\Re\bqty{R_\gamma + R_\beta} + \Im\bqty{R_\gamma - R_\beta}  }.
    \label{ft}
\end{equation}
For sufficiently small time intervals $t_1$ and $t_3$, the real part of this Fourier transform is exactly proportional to $J(\omega)(2N(\omega)+1)$. This follows from Taylor expanding Eq.~\eqref{fw}:
\begin{equation}
    f_{\pm\omega}(t) = -i t + \mathcal{O}(t^2).
\end{equation}

If $t_1$ and $t_3$ are sufficiently small compared to the exponents in $f_{\pm\omega}(t)$, Eq.~\eqref{ft} thus reduces to:
\begin{equation}
    \Re \bqty{\Tilde{R}(t_1,\omega_2,t_3)} \approx -\frac{\pi}{2} t_1 t_3 J(\omega_2)(2N(\omega_2)+1).
    \label{extr}
\end{equation}
Finally, we can divide by $-\frac{\pi}{2} t_1 t_3 (2N(\omega_2)+1)$ to obtain the spectral density by itself. Rather than directly setting $t_1$ and $t_3$ to a sufficiently small value, we evaluate Eq.~\eqref{extr} for a range of time intervals and fit the solution to a polynomial with variables $t_1$ and $t_3$. Then, for the term $a_{11} t_1 t_3$ in the polynomial, the coefficient $a_{11}$ directly corresponds to $-\frac{\pi}{2} J(\omega_2) (2N(\omega_2)+1)$. For the results discussed below, we extracted this coefficient from six time intervals, ranging from $0-0.05$~ps.

In an experimental setting, applying this protocol would thus have to involve the following steps: first, measure the real part of $R(t_0,t_2,t_0)$ for a range of waiting times and for a small $t_1,t_3=t_0$. Then apply a discrete Fourier transform to the measured signal, and finally select its real part.

\begin{figure}
\includegraphics[width=8.45cm]{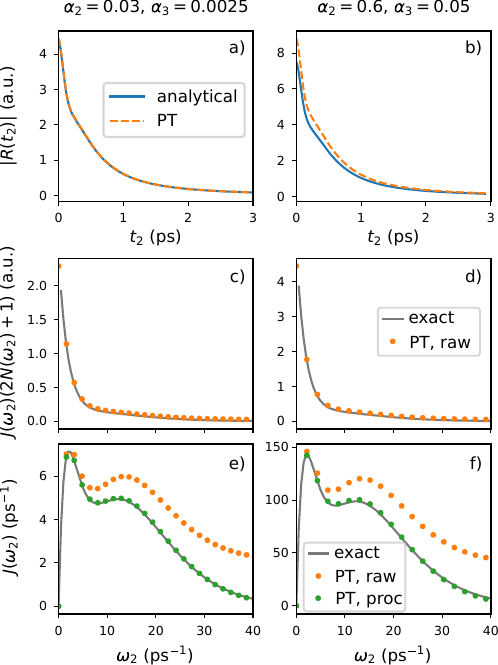}
\caption{Simulated response for the spectral density in Eq.~\eqref{sd2}, and two sets of system-bath coupling strengths $\alpha_{1,2}$. a-b) The modulus of the response in the time domain as a function of $t_2$ and $t_1=t_3=0.01$~ps, calculated with the analytical solution in Eq.~\eqref{r2} (blue) and PT-TEMPO (orange dashed). c-d) Extracted bath information computed with PT-TEMPO (orange dots), compared to the expected result (grey line). The offset has not been subtracted from the results, discussed in the main text. e-f) The extracted spectral density computed with PT-TEMPO and compared to the expected result (grey line). The orange dots represent the raw output without the offset subtracted; green dots correspond to the processed data.} 
\label{fig4}
\end{figure}

In Figs.~\ref{fig3} and~\ref{fig4}, we show the results of our extraction method for the spectral densities in Eqs.~\eqref{sd} and~\eqref{sd2} respectively. Panel a) in Fig.~\ref{fig3} shows the time domain signal corresponding to Eq.~\eqref{r2}, where $t_0=0.01$~ps. Here we have chosen to show the modulus of the signal, such that the phase envelope $e^{-i\epsilon(t_1 + t_3)}$ disappears. We find that for the chosen parameters ($\alpha_1=0.02$), the analytical (blue) and numerical (orange dashed) results agree well. 

We display the extracted bath information (orange dots) from the process tensor computation in panel b).  Broadly, the extracted data agrees well with the actual  $J_1(\omega)(2N(\omega)+1)$ (grey line). We do find however that the extracted curve does not completely decay to zero at large frequencies $\omega_2$, as expected. This becomes clearly visible in panel c), which displays the extracted spectral density $J_1(\omega)$. This is a consequence of the discrete nature of the numerical Fourier transform we are performing, which is discussed in Section 5 of the Supporting Information. Nonetheless, we can successfully extract the spectral density by removing the offset from zero in panel b), before dividing by $\frac{\pi}{2}(2N(\omega)+1)$ (green dots).

In panels d)-f) in Fig.~\eqref{fig3} we have increased the system bath coupling to $\alpha_1=0.4$. For increasing system-bath coupling strengths, we expect that our primary approximation in the analytical solution---limiting each interaction to a single quantum exchange---breaks down. The results in panel d) indeed confirm that the analytical solution does not agree as well with the numerically exact computation when $\alpha_1$ is increased. Panel f) shows that we can still successfully extract the spectral density from the numerical solution for this particular value of $\alpha_1$.

We repeat the extraction process for the spectral density $J_2(\omega)$ in Eq.~\eqref{sd2}, as presented in Fig.~\ref{fig4}. As shown in panels c) and f), our method is successful in resolving both peaks present in the spectral density.

To conclude, we have presented a method for extracting the spectral density of vibrational modes that couple to electronic transitions, using 2D electronic spectroscopy. The protocol would consist of sequentially probing two distinct electronic transitions as a function of the waiting time. This could for example be achieved by varying the polarisation of the pulses to probe states with orthogonal dipole moments, as is the case for the $Q_x$ and $Q_y$ states in chlorophylls~\cite{westenhoff:2012, schlau:2012, song:2019}. An alternative method for selecting different states could be to vary the color of the pulses~\cite{prall:2004}. By keeping the first and final time delay sufficiently small, the measured signal in the Fourier domain is directly proportional to the spectral density. We simulated this pulse sequence for a model chromophore coupled to a vibrational environment with a process tensor method that can compute multi-time correlations in a numerically exact way. To gain physical insight into the evolution of the system and environment during the pulse sequence, we compared the numerical results to an approximate analytical derivation. These results show that both an Ohmic spectral density and one with additional structure can be accurately extracted with our method.

We moreover found that the simulated response is zero under a standard Markovian master equation approach, indicating that non-Markovianity plays a significant role. The analytical derivation sheds more light on this: it shows that the response arises from the transfer of excitations, which leave and re-enter the system at a later time during the pulse sequence. Previous studies have suggested that an experimental witness of non-Markovianity should involve the measurement of multi-time correlations~\cite{milz:2019, fuxthesis}. Additional work has shown how a 2D NMR spectroscopy setup could be used to witness initial correlations by sequentially disentangling en entangling the system and environment~\cite{jatakia:2021}.
Beyond extracting spectral densities, the method introduced in this work could thus be an interesting step towards designing a protocol for a measure of non-Markovianity in open quantum systems.\newline

\textbf{Supporting Information:} Information on the computational resources and convergence parameters used for the PT-TEMPO calculations; a detailed description of the time evolution during $t_3$, the determination of the time-dependent coefficients, and of taking the weighted average over thermal states in the analytical derivation; further explanation on the error in the discrete Fourier transform of the data (PDF).

\begin{acknowledgments}
  R.\ d.\ W. acknowledges support from EPSRC (EP/W524505/1).  B.\ W.\ L.\ and J.\ K.\ acknowledge support from EPSRC (EP/T014032/1). A.\ W.\ C.\ wishes to acknowledge support from ANR Project ACCEPT (Grant No. ANR-19-CE24-0028).
\end{acknowledgments}

\bibliography{references}

\clearpage
\newpage
\renewcommand{\theequation}{S\arabic{equation}}
\renewcommand{\thefigure}{S\arabic{figure}}
\setcounter{section}{0}
\setcounter{equation}{0}
\setcounter{figure}{0}
\setcounter{page}{1}

\onecolumngrid
\section*{Supporting Information: Extracting Coupling-Mode Spectral Densities with Two-Dimensional Electronic Spectroscopy}
\twocolumngrid

\section{Computational resources}
The numerical computations discussed in the main text were performed with the PT-TEMPO method, which is part of the open source package OQuPy~\cite{oqupaper, oqupy}. The precision and efficiency of this method relies on the following three parameters: the time step $\delta t$, the maximum bath memory length $\Delta K_{max}$ and the maximal relative error in the singular value cutoff $\epsilon_{rel}$. The meaning and function of these parameters are discussed in detail in Ref.~\cite{oqupaper}. For this work, the computations were performed with $\delta t = 0.01$~ps, $\Delta K_{max} = 1000$ and $\epsilon_{rel}=10^{-7}$. We show in Ref.~\cite{dewit:2025} that these parameters produce numerically converged results. 

For the results shown in the main text, the computational cost depends on the system-bath coupling strength $\alpha$ and the particular form of the spectral density $J(\omega)$. Generally, stronger coupling and more structured spectral densities are computationally more expensive~\cite{oqupaper}. For an Ohmic spectral density and $\alpha=0.02$ in Fig.~\ref{fig3}a), it took 78 mins to compute the four-time correlation function $R(t_2; t_1=t_3=0.01$~ps$)$ for 300 time steps on a single core of a Intel i5 (8th Generation) processor, with a pre-computed process tensor. The construction of the process tensor took 43 mins. In contrast, the computation for the structured spectral density with $\alpha_2=0.6$ in Fig.~\ref{fig4}b) took 17 core hours for 300 time steps, plus 16 core hours to construct the process tensor.

\section{Evolution during the final interval}
Picking up from Eq.~\eqref{t2} in the main text, we apply $\hat{V}_1$ to excite the state into the $\ket{1}$ manifold at $\tau_3$:
\begin{multline}
    e^{-iE_{\vb{n}}(t_1 + t_2)  }\sum_i\Big[\beta_{i}(t_1) e^{-i\omega_it_2} \ket{1, \vb{n} + \vb{e}_i }
    \\+\gamma_{i}(t_1) e^{+i\omega_it_2} \ket{1, \vb{n} - \vb{e}_i  }\Big].
\end{multline}
During $t_3$, both terms will evolve back into a superposition in both excited state manifolds:
\begin{multline}
    \ket{1, \vb{n} \pm \vb{e}_i (t_3)} \approx  e^{-iE_{\vb{n}}t_3  }\alpha^\pm_i(t_3) \ket{1, \vb{n} \pm \vb{e}_i} \\
    + e^{-iE_{\vb{n}}t_3  }\sum_j \Big[\beta_{ij}^\pm (t_3) \ket{2, \vb{n} \pm \vb{e}_i +\vb{e}_j} \\
    + \gamma_{ij}^\pm (t_3) \ket{2, \vb{n} \pm \vb{e}_i -\vb{e}_j}     \Big],
    \label{firstterm}
\end{multline}
where $\alpha_i^\pm$, $\beta_{ij}^\pm$, $\gamma_{ij}^\pm$ are now new coefficients that will have similar forms to $\alpha$, $\beta_i$ and $\gamma_i$ respectively. 

After $t_3$, we apply the final operator $\hat{V}_2$ and take the trace. Since we are applying all four dipole operators to the ket-side of the density matrix, the bra-side will still be in the ground state after $t_3$, {\it i.e.} $\bra{0, \vb{n}}$. To obtain a non-zero trace, the only terms in the sum over $i$ and $j$ that contribute to the final solution are those for which $i=j$ and such that the final number state is $\vb{n}$.
For example:
\begin{equation}
    \Tr\sum_{ij}\beta^-_{ij}(t_3)\ket{0, \vb{n} - \vb{e}_i +\vb{e}_j   }\bra{0, \vb{n}} = \sum_{ij}\delta_{ij}\beta^-_{ij}
    \label{trace}
\end{equation}
Thus, in Eq.~\eqref{firstterm}, only $\gamma^+_{ii}$ and $\beta^-_{ii}$ contribute to the final solution. After taking the trace, the correlation function will then be of the form:
\begin{equation}
    \sum_i \bqty{\beta_{i}(t_1) e^{-i\omega_{ii} t_2} \gamma^+_{ii}(t_3) + \gamma_{i}(t_1) e^{i\omega_{ii} t_2}\beta^-_{ii}(t_3)},
    \label{sol0}
\end{equation}
We discuss the determination of the coefficients in the above expression in the next Section.

\section{Weisskopf-Wigner approach}
Here we describe in detail the derivation of the coefficients in Eq.~\eqref{sol0}, focussing on the first time interval $t_1$. First, we insert the coefficients $\alpha(t_1), \beta_i(t_1), \gamma_i(t_1)$ into the time-dependent Schrödinger equation:
\begin{equation}
    i\dv{\alpha(t_1)}{t} = E_2\alpha(t_1) + \sum_i \bqty{g_i^*\sqrt{n_i+1}\beta_i(t_1) + g_i\sqrt{n_i}\gamma_i(t_1)},
    \label{eom0}
\end{equation}
\begin{multline}
    i\dv{\beta_i(t_1)}{t} = \pqty{ E_1 + \omega_i}\beta_i(t_1) + g_i\sqrt{n_i+1}\alpha(t_1)  \\
    + \sum_m \bqty{g_m^*\sqrt{n_m+1}b^+_{im}(t_1) + g_m\sqrt{n_m}c^+_{im}(t_1)},
    \label{eom1}
\end{multline}
\begin{multline}
    i\dv{\gamma_i(t_1)}{t} = \pqty{ E_1 - \omega_i}\gamma_i(t_1) +  g_i^*\sqrt{n_i}\alpha(t_1) \\
    + \sum_m \bqty{g_m^*\sqrt{n_m+1}b^-_{im}(t_1) + g_m\sqrt{n_m}c^-_{im}(t_1)},
    \label{eom2}
\end{multline}
 where $E_{1,2}$ are the eigenvalues of $\hat{H}_S$ in Eq.~\eqref{ham} and $g_i$ the system-bath coupling of mode $i$ in $\hat{H}_I$. $n_i$ refers to the occupation of mode $i$. In order to ensure dissipation of the same order in $g_i$ as in the evolution of $\alpha(t_1)$, the equations of motion for $\beta_i(t_1)$ and $\gamma_i(t_1)$ additionally include coupling terms to states involving bath modes $m$, which evolve according to the coefficients $b^{\pm}_{im}(t_1)$ and $c^{\pm}_{im}(t_1)$:
\begin{flalign}
    &i\dv{b^{+}_{im}(t_1)}{t} = \pqty{E_2 + \omega_i + \omega_m}b^{+}_{im}(t_1) + g_m\sqrt{n_m+1}\beta_i(t_1) \\
    &i\dv{c^{+}_{im}(t_1)}{t} = \pqty{E_2 + \omega_i - \omega_m}c^{+}_{im}(t_1) + g_m^*\sqrt{n_m}\beta_i(t_1) \\
    &i\dv{b^{-}_{im}(t_1)}{t} = \pqty{E_2 - \omega_i + \omega_m}b^{-}_{im}(t_1) + g_m\sqrt{n_m+1}\gamma_i(t_1) \\
    &i\dv{c^{-}_{im}(t_1)}{t} = \pqty{E_2 - \omega_i - \omega_m}c^{-}_{im}(t_1) + g_m^*\sqrt{n_m}\gamma_i(t_1).
\end{flalign}
Now, following a Weisskopf-Wigner approach~\cite{weisskopf:1930, breuer:2002}, we first solve the equations of motion for $b^\pm_{im}(t_1)$ and $c^\pm_{im}(t_1)$ in terms of $\beta_i(t_1)$ and $\gamma_i(t_1)$. For example:
\begin{flalign}
    b^+_{im}(t_1) &= -ig_m\sqrt{n_m+1}\int_0^t \dd t^\prime e^{-i\pqty{E_2 +   \omega_i + \omega_m}(t-t^\prime)}\beta_i(t^\prime)\\
    c^+_{im}(t_1) &= -ig_m^*\sqrt{n_m}\int_0^t \dd t^\prime e^{-i\pqty{E_2  + \omega_i -\omega_m}(t-t^\prime)}\beta_i(t^\prime),
\end{flalign}
and similar for $d^-_{im}(t_1)$ and $c^-_{im}(t_1)$. We can now insert these back into the equations of motion for $\beta_i(t_1)$ and $\gamma_i(t_1)$. We will explicitly show the next steps for $\beta_i(t_1)$. Inserting $b^+_{im}(t_1)$ and $c^+_{im}(t_1)$:
\begin{multline}
     \dv{\beta_i(t_1)}{t} = -i(E_1 + \omega_i)\beta_i - i g_i \sqrt{n_i+1} \alpha(t_1) \\- \int_0^t \dd t^\prime f_m^+(t-t)^\prime)e^{-\omega_i(t-t^\prime)}\beta_i(t^\prime),
     \label{eom3}
 \end{multline}
 where 
 \begin{equation}
     f_m^+(\tau) = \sum_m \abs{g_m}^2 e^{-E_2\tau}\big[(n_m+1)e^{-i\omega_m\tau}
     +n_m e^{i\omega_m\tau}\big].
 \end{equation}
In Eq.~\eqref{eom3}, the complex exponential dependent on $\omega_i$ has been explicitly separated from $f_m^+(t-t^\prime)$ to emphasize that this term will cancel in the next step. In order to solve for $\beta_i(t_1)$, we first move into the interaction picture $\beta_i(t_1)=\Tilde{\beta_i}(t_1)e^{-i( E_1+\omega_i)t}$. For simplicity, we assume the influence of the bath modes $m$ on the time evolution of $\beta_i(t_1)$ is Markovian~\footnote{ 
Note that while we make this Markovian approximation for the dissipation acting on $\beta_i(t_1)$, our overall treatment is not Markovian, in that we rely on system-environment entanglement.}, though we note that an exact solution could be found via Laplace transforms~\cite{breuer:2002}.
We then obtain:
\begin{equation}
     \dv{\Tilde{\beta_i}(t_1)}{t} =  -i g_i\sqrt{n_i+1}e^{i(E_1+\omega_i)t}\alpha(t_1)-\Lambda_1 \Tilde{\beta_i}(t_1).
\end{equation}
with
\begin{equation}
     \Lambda_1 = \int_0^\infty \dd \tau^\prime f_m^+(\tau) e^{i E_1\tau}.
\end{equation}
$\Lambda_1$ is composed of a real part $\Gamma_1$ leading to exponential decay and an imaginary part corresponding to the negative Lamb shift $S(-2\Omega)$:
\begin{flalign}
    &\Gamma_1 = \pi N\pqty{2\Omega}J(2\Omega);\label{gammamin}\\
    &S(\mu) = \mathcal{P} \int^\infty_{-\infty} \frac{J(\omega)\pqty{N(\omega)+1} +J(-\omega)N(-\omega)}{\mu - \omega}.
    \end{flalign}
We can follow exactly the same procedure for $\gamma_i(t_1)$, which leads to the same dissipation term. Thus, for the equations of motion for $\beta_i(t_1)$ and $\gamma_i(t_1)$ we now have:
\begin{flalign}
    &i\dv{\beta_i(t_1)}{t} = \pqty{ E_1 + \omega_i-i\Lambda_1}\beta_i(t_1) + g_i\sqrt{n_i+1}\alpha(t_1) \label{eombeta}\\
    &i\dv{\gamma_i(t_1)}{t} = \pqty{ E_1 - \omega_i-i\Lambda_1}\gamma_i(t_1) + g_i^*\sqrt{n_i}\alpha(t_1). 
    \label{eomgamma}
\end{flalign}

We now repeat the same Weisskopf-Wigner procedure on the differential equation for $\alpha(t_1)$: we solve Eqs.~\eqref{eombeta} and~\eqref{eomgamma} in terms of $\alpha(t_1)$, and insert $\beta_i(t_1)$ and $\gamma_i(t_1)$ into Eq.~\eqref{eom0}. By moving into the interaction picture, making a Markov approximation and solving the differential equation we obtain:
\begin{equation}
    \alpha(t_1) = e^{-i(E_2 + S(2\Omega))t-\Gamma_2 t},
\end{equation}
where $S(2\Omega)$ now corresponds to a positive Lamb shift with a dissipation rate $\Gamma_2 = \pi \pqty{N\pqty{2\Omega}+1}J(2\Omega)$.
Finally, substituting $\alpha(t_1)$ into the Eqs.~\eqref{eombeta} and~\eqref{eomgamma} gives us the final expressions for $\beta_i(t_1)$ and $\gamma_i(t_1)$:
\begin{flalign}
    &\beta_i(t_1) = g_i\sqrt{n_i+1}\pqty{\frac{e^{-i(E_1^\prime+\omega_i) t_1-\Gamma_1t_1}-e^{-iE_2^\prime t_1-\Gamma_2t_1}} {E_2^\prime - E_1^\prime - \omega_i - i\Gamma_2 + i\Gamma_1}}, \label{c1}\\
    &\gamma_i(t_1) =  g_i^*\sqrt{n_i}\pqty{\frac{e^{-i\pqty{E_1^\prime -\omega_i}t_1-\Gamma_1 t_1}-e^{-iE_2^\prime t_1 - \Gamma_2 t_1} }{E_2^\prime - E_1^\prime + \omega_i - i\Gamma_2+i\Gamma_1}},
    \label{c2}
\end{flalign}
where the Lamb shifts have been absorbed into the system eigenvalues, {\it i.e.} $E_1^\prime = E_1 + S(-2\Omega)$ and $E_2^\prime = E_2 + S(2\Omega)$.

We now repeat the entire procedure for the $t_3$-dependent coefficients, keeping in mind that only the coefficients with indices $i=j$ contribute to the final solution, as shown in Eq.~\eqref{trace}. This results in the following solutions:
\begin{flalign}
    &\gamma_{ii}^+(t_3) = g_i^*\sqrt{n_i+1}\pqty{\frac{e^{-i(E_1^\prime+\omega_i) t_3-\Gamma_1t_3}-e^{-iE_2^\prime t_3-\Gamma_2t_3}} {E_2^\prime - E_1^\prime - \omega_i - i\Gamma_2 + i\Gamma_1}},\label{c3}\\
    &\beta_{ii}^-(t_3) =  g_i\sqrt{n_i}\pqty{\frac{e^{-i\pqty{E_1^\prime -\omega_i}t_3-\Gamma_1 t_3}-e^{-iE_2^\prime t_3 - \Gamma_2 t_3} }{E_2^\prime - E_1^\prime + \omega_i - i\Gamma_2+i\Gamma_1}}.
    \label{c4}
\end{flalign}

\section{Obtaining the final solution}
We can now insert the solutions for the coefficients in Eqs.~\eqref{c1}-\eqref{c4} into Eq.~\eqref{sol0}. To obtain the final solution for the signal $R$ stated in the main text, we lastly take the thermal average over all vibrational configurations. Since the solution is linear in configurations, this corresponds to a weighted sum over occupation vectors $\vb{n}$:
\begin{equation}
    R = \sum_{\vb{n},i}P_{\vb{n}}\bqty{\beta_{i}(t_1) e^{-i\omega_{ii} t_2} \gamma^+_{ii}(t_3) + \gamma_{i}(t_1) e^{i\omega_{ii} t_2}\beta^-_{ii}(t_3)},
\end{equation}
where $P_{\vb{n}}$ is the Boltzmann-weighted probability of $\vb{n}$, stated in the main text. Additionally, since the environment is composed of a continuum of modes, we can replace the sum over modes $i$ by an integral over vibrational frequencies, resulting in the spectral density dependence. For example:
\begin{equation}
    \sum_{\vb{n},i}P_{\vb{n}}\, \abs{g_i}^2 (n_i+1) = \int_0^\infty\dd \omega J(\omega) (N(\omega)+1),
\end{equation}
using the definition of the spectral density, $J(\omega)= \sum_i \abs{g_i}^2 \delta(\omega - \omega_i)$, and the Bose-Einstein occupation number $N(\omega)$. This results in the final solution of $R$, described in the main text in Eqs.~\eqref{r2}-\eqref{fw}.

\section{Error in Fourier transform as function of time step}
\begin{figure}
\includegraphics[width=8.45cm]{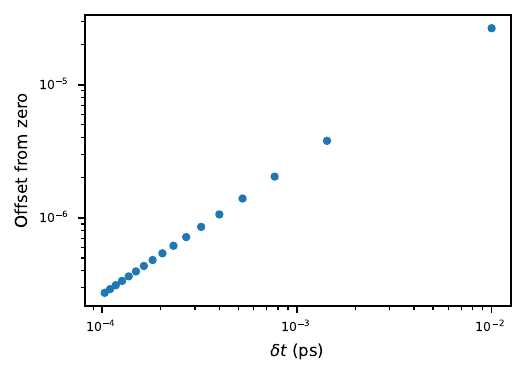}
\caption{Magnitude of the offset from zero of $\mathrm{Re}\bqty{\Tilde{R}(t_1,\omega_2,t_3)}$, when $\omega_2\rightarrow\infty$, for different values of the time step $\delta t$. Data points were computed using the Ohmic spectral density $J_(\omega)$ with $\alpha=0.4$.}  
\label{S1}
\end{figure}

In the main text, we mentioned that the extracted data has a finite offset from zero at large frequencies compared to the exact form $J(\omega_2)(2N(\omega_2)+1)$, which decays to zero. In Fig.~\ref{S1} we show the magnitude of this offset, calculated by performing a discrete Fourier transform on the analytical solution to $\mathrm{Re}\bqty{R(t_1,t_2,t_3)}$ in Eqs.~\eqref{r2} and~\eqref{FT}, for different values of the time step $\delta t$. The largest time step displayed, $\delta t = 0.01$~ps, corresponds to the value used in the main results in Figs.~\ref{fig3} and~\ref{fig4}. By extrapolation, we can see that the offset tends to zero for an infinitesimally small time step. Since both the analytical and numerical PT-TEMPO results decay to a finite value rather than to zero, we can conclude that this error is likely a consequence of the discrete nature of the numerical Fourier transform. Because the time step sets the finite length of the frequency window, the signal may be distorted due to aliasing effects, which cause replicas of the signal in the Fourier domain to overlap. 

Thus we find that, if it were computationally feasible to perform the PT-TEMPO calculation with a smaller time step, the offset after performing the discrete Fourier transform would disappear.

\end{document}